\documentclass[twocolumn,showpacs,preprintnumbers,amsmath,amssymb]{revtex4}

\usepackage{graphicx}
\usepackage{epsfig}
\usepackage{longtable}
\usepackage{longtable}

\begin{document}

\title{Sizes and shapes of very heavy nuclei in \mbox{high-$K$} states}

\author{M.~Palczewski$^{1}$}
\author{P.~Jachimowicz$^{2}$}
\author{M.~Kowal$^{1}$} \email{michal.kowal@ncbj.gov.pl}

\affiliation{$^1$ National Centre for Nuclear Research, Pasteura 7, 02-093 Warsaw, Poland}
\affiliation{$^2$ Institute of Physics, University of Zielona G\'{o}ra, Z. Szafrana 4a, 65-516 Zielona G\'{o}ra, Poland}

\date{\today}

\begin{abstract}
We have investigated shapes and sizes of selected two- and four-quasiparticle
\mbox{high-$K$} states in nobelium and ruthefordium isotopes within
 the microscopic-macroscopic model with the deformed Woods-Saxon potential.
Excited nuclear configurations were obtained by blocking single-particle states
 lying close to the Fermi level.
 Their energies and deformations were found by the four-dimensional energy
 minimization over shape variables.
 We have selected the most promising candidates for \mbox{$K$-isomers}
 by analyzing the isotopic dependence of excitation energies, and compared
 our results to available experimental data.
 We calculated differences in quadrupole moments and charge radii
 between nuclei in their \mbox{high-$K$} and ground states and
 found their quite different pattern for four-quasiparticle states in
 neighbouring No and Rf isotopes.
 The leading role of the quadrupole and hexadecapole deformations as well as
 the importance of higher rank symmetries is also discussed.
 The current development of laser techniques and the resulting ability to
 measure discussed effects in near future are the motivation of
 our study.

\end{abstract}


\maketitle

\section{INTRODUCTION}

 Shapes and sizes of atomic nuclei are their primary characteristics.
 Their study acquires a flavour of novelty for very heavy nuclear systems
 for which still relatively little is known.
 Quite recently, a development of laser spectroscopy methods,
 improving measurements of hyperfine splitting and isotope shifts
\cite{Kluge2003,Gangrsky2006,Backea2015,Campell2016,Lat2016,Bloch2020}, has
 been achieved.
 The laser-based spectroscopic techniques allowed to perform
 model-independent and direct measurement of basic quantities related
 to nuclear shapes: electric quadrupole moments and mean-square charge radii,
 for very heavy nuclei.
 Those measurements provide not only important and new information on nuclear
 ahapes and sizes but also provide a test for theoretical models as they are
 sensitive not only to the bulk properties but also to the single-particle
 spectrum/configurations.
 Quite recently, such laser-based data on ground-states
 of $^{252,253,254}$No were discussed and published by Raeder et al.
 \cite{Raeder2018}. Nobelium is so far the heaviest element in which
 these quantities could be measured.

 An interesting question is what are sizes and shapes of the
 \mbox{high-$K$} isomers which frequently occur in these nuclei.
 In particular, does a multi-quasiparticle excitation increase or decrease
 the size of a nuclear system relative to that of the ground state?
 A pioneering experiment in which the collinear laser spectroscopy was applied
 to the four-quasiparticle (qp) isomeric state in $^{178}Hf^{m2}$ pointed out
 that the change in nuclear mean-square charge radius
 is $\delta \langle r^{2} \rangle=-0.059(9)fm^{2}$ \cite{Boos1994}. This means
 that nuclear size of the isomer is significantly smaller than that of the
 ground state.
 How the shape of very heavy multi-quasiparticle isomers changes along the
 isotopic chains is another interesting question.
 Motivated by the mentioned unprecedented development in accuracy of the laser
 measuring techniques, here we would like to provide some answers to those
 questions.

A review of the current experimental knowledge on isomers in the heaviest
 nuclei can be found in \cite{Ackermann2015,Asai2015,Theisen2015,Dracoulis2016, Walker2020}, while theoretical overview based on the
Nilsson - Strutinsky approach was given in \cite{Walker2016}.
As follows from the cited works, for nuclei close to nobelium and
rutherfordium the occurrence of \mbox{high-$K$}, low-lying isomeric states
 seems to be very likely due to the predicted existence of the deformed shell gaps at $Z=100$ and $N=152$ \cite{Greenlees2008}.

There is another extremely interesting aspect of such research:
the size of a nucleus, whether in the ground or isomeric state, is closely
 related to physics of the alpha decay.
This is of particular importance in the context of a search for hindrance
 mechanisms in this decay.
 Recently, we predicted \cite{Jachimowicz2018} a quite strong hindrance against
 alpha decay for four-qp states: $K^{\pi} = 20^{+}$ and/or $19^{+}$,
 and two-quasiproton state: $K^{\pi} = 10^{-}$ in darmstadium nuclei
 and for some odd and odd-add superheavy nuclei \cite{Jachimowicz2015}.
 Together with their relatively low excitation, this suggests a possibility
 that they could be isomers with an extra stability.

Although a link of such quantities as the alpha-decay energy
 $Q_\alpha$ and half-live with the size and shape of the nucleus is not always
 obvious, intuitively one may expect some relationships.
i) The size of a nucleus will have effect on the pre-formation probability
 of the alpha cluster inside nucleus.
The probability of alpha particle formation should be different in its
 periphery.
ii) The action integral used to estimate alpha half-life is strongly dependent
 on the turning points which are in turn very much dependent on the nuclear
 radius.
iii) For an isomer, $Q_\alpha$ energy available in the emission process may
  change significantly, and that will, in a non-direct way, affect rather
 considerably the integration limits (see previous point),
 and so the whole tunneling process through such effectively changed potential
 barrier.

 The main goal of this paper is to provide predictions for sizes and shapes
 of very heavy nuclei in \mbox{high-$K$} states.
 Relevant quantities: electric quadrupole moments and nuclear charge radii
 are calculated within the MM model.
 Since s.p. wave functions in such a model are used only to calculate shell
 corrections, they are not suited for the quantities of interest for this
 work. Therefore we use the liquid-drop model formulas which allow
 to calculate both, the quadrupole moment and the mean-square radius
 as the functions of only deformations.
 It has to be emphasized, that this approach seems a approximation
 and its usefulness for high-K configuration will depend on how well it
 compares with experimental data.

\section{THE METHOD}

To obtain energies and deformations for the ground (gs) and excited (ex) states
 the microscopic-macroscopic (MM) method is used. In the frame of this approach microscopic
energy is calculated via applying the Strutinski shell and pairing
correction \cite{Strutinski1966_67} method to single particle levels
 of the deformed Woods-Saxon potential \cite{Cwiok1987}.
The $n_{p}=450$ lowest proton levels and $n_{n}=550$ lowest neutron levels
 from the $N_{max}=19$ lowest shells of the deformed harmonic oscillator
 are taken into account in the diagonalization procedure.
 Standard values of $\hbar\omega_{0}=41/A^{1/3}~MeV$  for the oscillator
 energy and $\gamma=1.2 \hbar\omega_{0}$ for the Strutinski smearing parameter
 $\gamma$, and the sixth-order correction polynomial are used in the
 calculation of the shell correction.
For the macroscopic part we used the Yukawa plus exponential model
 \cite{Krappe1979} with parameters specified in \cite{Muntian2001}.
 Thus, all parameter values are kept exactly the same as in all recent
 applications of the model to heavy and superheavy nuclei, e.g.,
\cite{Kowal2010,Kowal2010_2,archKowal2012,Kowal2012,Jachimowicz2012_20,Jachimowicz2013,Jachimowicz2014,Jachimowicz2017,Jachimowicz2017_2,Jachimowicz2018}.

 In the considered region of nobelium nuclei, the ground- as well as excited
 states are expected to be well deformed and axially- and reflection-symmetric.
 This means that intrinsic parity of states is well defined and
 that $K$ is a "good" quantum number.
 Admittedly, this assumption cannot be exact for \mbox{high-$K$} states,
in which the time-reversal breaking effects are expected to break the axial
 symmetry to some degree.

In the present study we used four deformation parameters within the
  standard $\beta$ parametrization.
The nuclear radius vector $R(\theta ,\phi)$ is parameterized via spherical
harmonics ${\rm Y}_{\lambda \mu}(\theta,\phi)$ as follows:
\begin{equation}
\label{radius}
R(\theta)= c R_0 [ 1+\sum_{\lambda=2,4,6,8}\beta_{\lambda 0}
{\rm Y}_{\lambda 0}(\theta)],
\end{equation}
where $c$ is the volume-fixing factor depending on deformation and $R_0$ is
 the radius of a spherical nucleus taken as: $R_0=1.16 \cdot A^{1/3} fm$.

To find the gs minima, i.e. energies and shapes,
the four-dimensional energy minimization over $\beta_{20}-\beta_{80}$ is
 performed using the gradient method. Such minimization is repeated dozens
 of times for a given nucleus with different starting values of deformations.
To obtain excitation energies, after blocking of a chosen configuration,
 a similar minimization procedure, over the same deformations, was performed
 once again.

 We would like to emphasize that the used MM model, among all existing, offers
 not only an excellent agreement with existing data in the region of
 super-heavy nuclei (concerning: masses, deformations \cite{Kowal2010},
 $Q_\alpha$-energies \cite{Jachimowicz2014}, first and second fission
barriers in actinides \cite{Jachimowicz2012_20}, etc.) but also, what is even
 more important for spectroscopic studies,
provides at the same time two prominent enough energetic shell gaps
at $Z=100$ in protons and $N=152$ in neutrons. 
Without this, it seems, no realistic predictions for \mbox{$K$-isomers} 
in this region of nuclei were possible.

\section{RESULTS AND DISCUSSION}

\subsection{Excitation energies of the selected 2- and 4-qp \mbox{high-$K$},
 possibly isomeric states}

Promising candidates for metastable isomers are
energetically low-lying \mbox{high-$K$} states. The natural candidates are
 configurations built by particles blocked on levels in close proximity to
 the Fermi energy.
In the region of No and Rf nuclei such characteristic multi-quasiparticle
configurations are as follows:

\begin{itemize}
\item two-proton (2-qp isomer):

$ K^{\pi} = \pi^2 8^{-} \: \{\pi7/2^{-}[514] \otimes \pi9/2^{+}[624]\}$,
\item two-neutron (2-qp isomer):

$ K^{\pi} = \nu^2 8^{-} \: \{\nu7/2^{+}[624] \otimes \nu9/2^{-}[734]\}$,
\item two-proton plus two-neutron (4-qp isomer), build on two configurations above:

$K^{\pi}=\pi^2 \nu^2 16^{+}\: \{\pi^2 8^{-} \otimes \: \nu^2 8^{-}\}$.

\end{itemize}

\begin{figure}[t]
\centerline{\includegraphics[scale=0.7]{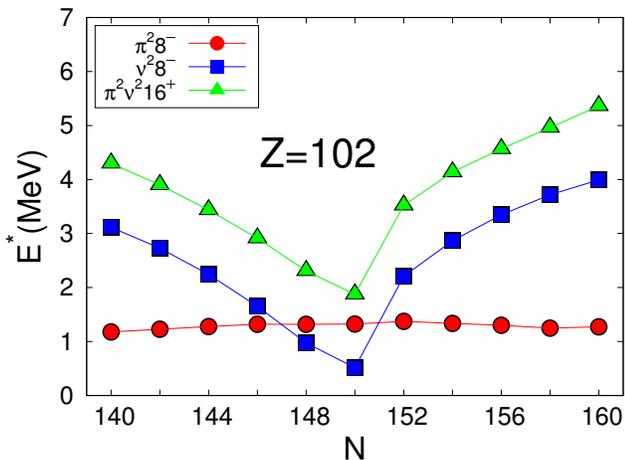}}
\caption{{\protect
Calculated excitation energies (\ref{eprime}) for multi-qp \mbox{high-$K$} states in No isotopes. Configurations:
$\pi^2 8^{-}=\{\pi7/2^{-}[514] \otimes \pi9/2^{+}[624]\}$,
$\nu^2 8^{-}=\{\nu7/2^{+}[624] \otimes \nu9/2^{-}[734]\}$,
$\pi^2 \nu^2 16^{+}=\{\pi^2 8^{-} \otimes \: \nu^2 8^{-}\}$.
The lines are drown to guide an eye.}}
\label{Ex102}
\end{figure}

\begin{figure}[t]
\centerline{\includegraphics[scale=0.7]{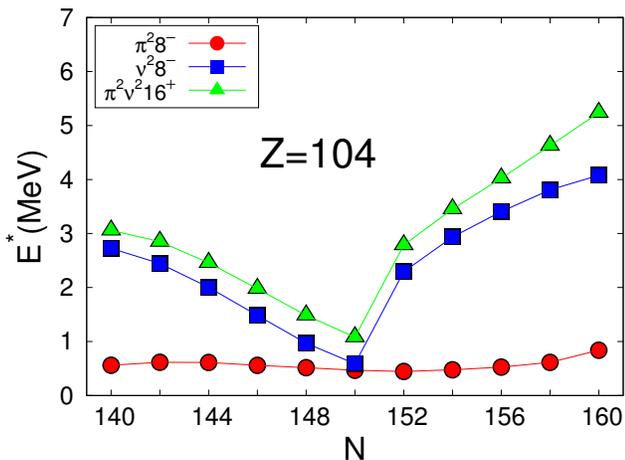}}
\caption{{\protect The same as in Fig. \ref{Ex102}, but in Rf isotopes.}}
\label{Ex104}
\end{figure}

They will be the subject of our further considerations concerning
their sizes and shapes.
However, first we discuss their excitation energies.
As explained earlier, for a given nucleus, from independently performed
 minimizations, we obtained deformations and energies of ground states and
 the constrained minima for selected few-qp configurations.
 The energy difference between these minima gives the excitation energy of
 a certain multi-qp configuration:
\begin{equation}
\label{eprime}
E^{*} =\Delta E=E^{ex}-E^{gs}.
\end{equation}
Those energies are shown for isotopic chains of No in Fig.~\ref{Ex102},
and Rf, in Fig.~\ref{Ex104}.
As can be seen, excitation energies of two-proton qp state (red dots) are
very low-lying for both elements,
what makes them promising candidates for 2-quasiproton isomers in the
whole range of considered neutron numbers $N$.
One can also see that the extremely low $E^{*}$ values for
 $\pi^2 8^{-}$ state
are obtained in Rf isotopes, and that in both, No and Rf elements,
excitation energies of such 2-quasiproton configuration
show very weak isotopic dependence.

From these figures one can choose the best candidates
for two-quasineutron isomers (blue dots).
As we clearly see for both considered elements, the lowest excitation
energies of the $\nu^2 8^{-}$ configuration occur for $N=150$.
Moreover, in $^{250}$No (Fig.~\ref{Ex102}), such two-quasineutron
state lies much lower than the two-quasiproton one,
while in $^{248,250}$No as well as in $^{254}$Rf both corresponding
 $E^{*}$ values are roughly similar.
In all remaining cases, the proton 2-qp configuration
always lies below the 2-quasineutron state.

 The $N$-dependence of these two (proton and neutron) energies allows
 a prediction of the most favorable four-quasiparticle $\pi^2 \nu^2 16^{+}$
 candidates. As can be seen in Figs.~\ref{Ex102} and \ref{Ex104} in green,
such 4-qp \mbox{high-$K$} isomeric state can appear most likely
in $^{252}$No and $^{254}$Rf because of the smallest excitation energy.

\begin{figure}[h]
\centerline{\includegraphics[scale=0.7]{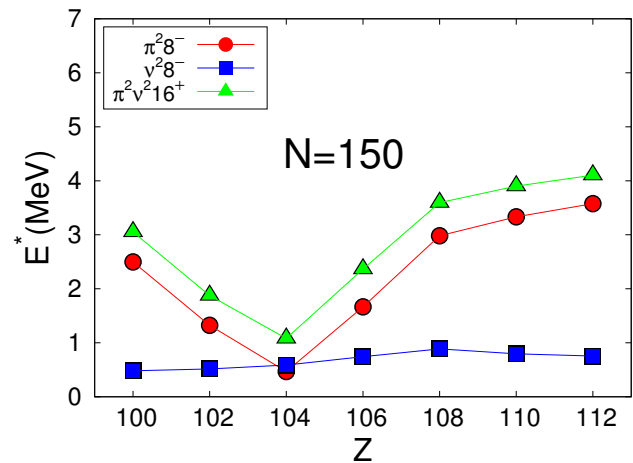}}
\caption{{\protect The same as in Fig.~\ref{Ex102}, but for $N=150$ isotones in the range from fermium to copernicium.}}
\label{Ex150}
\end{figure}

\begin{figure}[h]
\centerline{\includegraphics[scale=0.7]{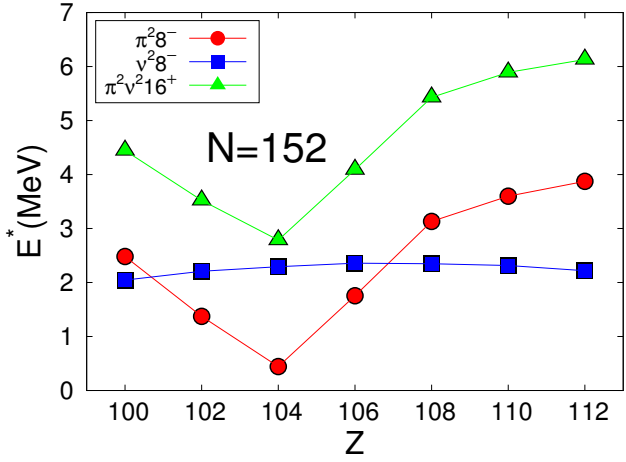}}
\caption{{\protect The same as in Fig.~\ref{Ex102}, but for $N=152$ isotones in the range from fermium to copernicium.}}
\label{Ex152}
\end{figure}

It seems interesting to make a similar analysis for specific neutron shells $N=150$ and $152$
for different elements from this area. Such isotonic chains for $Z=100-112$ are shown
in Figs.~\ref{Ex150} and \ref{Ex152} for $N=152$ and $N=152$, respectively.
This time we see very low lying two-neutron
configuration $\nu^2 8^{-}$ practically for all taken nuclei with neutron number $N=150$.
 This confirms the previous observation that the best candidate
 isomers is $^{254}$Rf.
Excitation of two-neutrons from the closed $N=152$ shell costs more energy
 as seen in Fig.~\ref{Ex152} in which this excitation is shown (in blue).
 This behavior of neutrons basically blocks the possibility
 for the creation of any 4-qp isomers for all nuclei possessing such number
 of neutrons.
 However, one can notice a very promising candidate for two-proton isomer
 in $^{256}$Rf.
 From Fig.~\ref{Ex152} one can not still completely exclude possible formation
 of the two-proton isomer in $^{254}$No and $^{258}$Sg. Similar conclusion
 about the two-proton state can be drown
 from Fig.~\ref{Ex150} and concerns $^{252}$No and $^{256}$Sg.

\subsection{Obtained excitation energies vs experimental data}

Before discussing shapes let us discuss the compliance of the obtained
 $E^{*}$ values with experimental data. As already mentioned, those
 configurations are relatively well-known
 for \mbox{$K$-isomers} in nobelium and rutherfordium isotopes investigated
 here (see the data set collected in \mbox{Table 2} in \cite{Theisen2015}).

\begin{figure}[h]
\centerline{\includegraphics[scale=0.45]{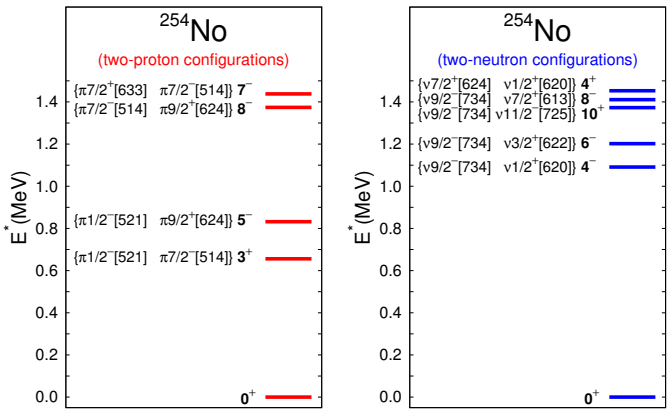}}
\caption{{\protect Calculations of $^{254}$No two-quasiproton states (left) and
two-quasineutron states (right) with excitation energy $E^{*} \leq 1.5~MeV$.}}
\label{254No}
\end{figure}

We start our comparison with $^{254}$No for which the first observations of excited state postulated as the
isomeric one were carried out already almost 50 years ago \cite{Ghiorso1973}.
Observed band characterised by $E^{*}=0.988~MeV$ \cite{Tandel2006} was assigned
there to the two-quasiproton $K^{\pi}=3^{+}$ state.
This and other experimental observations can be compared with our theoretical
results for $^{254}$No, shown in Fig.~\ref{254No}.

Let us note that in Fig.~\ref{254No} all possible combinations
of two-quasiparticle configurations for protons (red colour) and neutrons (blue colour)
that one can obtain from the deformed Woods-Saxon model after
4-dimensional minimization over all $\beta_{\lambda 0}$ in (\ref{radius}) and
 lying at energy lower than $E^{*} = 1.5~MeV$ has been shown.
One can see e.g. that the $3^{+}$ state
 $\{\pi 1/2^{-}[521] \otimes \pi 7/2^{-}[514]\}$
has the excitation energy a little below $0.7~MeV$ and it is the lowest
obtained possible two-quasiproton configuration.
There is also another low-lying proton state, not very distant
 ($\sim 200~keV$) from it, namely:
$K^{\pi}=5^{-}$ state: $\{\pi 1/2^{-}[521] \otimes \pi 9/2^{+}[624]\}$
and this one more closely matches the measured energy $E^{*}=0.988~keV$ \cite{Tandel2006}.

The difficulty in assigning the appropriate configuration is even more apparent
for the next important candidate for isomeric configuration, namely
 $K^{\pi}=8^{-}$, which is located around $E^{*} \approx 1.4~MeV$ in both
 proton and neutron spectra of most MM models.
Exact predictions of excitation energies
in various microscopic-macroscopic  approaches
for such two-quasiparticle \mbox{$K^{\pi} = 8^{-}$} states
are collected in \mbox{Table III} of \cite{Theisen2015}.
One can see that all of them are in the range $\langle 1.1  \div 1.5
\rangle~MeV$, while the exact
measured values of $E^{*}$ (in $keV$) are the following: 1293 \cite{Herzberg2006,HerzbergNAT2006},
1296 \cite{Tandel2006}, 1295(2) \cite{Hessberger2010}, 1297(2) \cite{Clark2010}.
In our calculations the following two-proton configuration:
$\{\pi 7/2^{-}[514] \otimes \pi 9/2^{+}[624]\}$ with $E^{*} \approx 1.4~MeV$
(left side of Fig.~\ref{254No}) may be assigned to the experimental excitation
 of this $K^{\pi}=8^{-}$ state.
On the other hand, looking on the right side of Fig.~\ref{254No}, where excitation
energies of two-neutron quasiparticle states are presented, one can notice another just as likely
$K^{\pi}=8^{-}$ configuration and attribute it as:
$\{\nu 9/2^{-}[734] \otimes \nu 7/2^{+}[613]\}$ at the same energy.
Interestingly, one can also see in our neutron spectrum another high spin state $K^{\pi} = 10^{-}$
near it, this time as a result of adding the following two-neutron
components: $\{\nu 9/2^{-}[734] \otimes \nu 11/2^{-}[725]\}$.

Second candidate which one can consult with the obtained values of excitation energies is $^{252}$No.
The experimental excitation energy for the \mbox{$K$-isomer} in this nucleus
  is $1.25~MeV$
\cite{Robinson2008,Sulignano2012} with the half-life $T_{1/2} = 109(6)~ms$.
According to Fig.~\ref{Ex102}, the state that best suits this energy is the
 two-quasiproton excitation
$\pi^2 8^{-}=\{\pi7/2^{-}[514] \otimes \pi9/2^{+}[624]\}$
 with $E^{*}$ about $1.3~MeV$ in our calculation.
Two neutron state considered earlier in $^{252}$No, namely: $\{\nu 9/2^{-}[734] \otimes \nu 7/2^{+}[613]\}$,
this time is much higher in energy ($ E^{*} = 1.97~MeV$).
However, as discussed in the previous section and shown in Fig.~\ref{Ex102},
there is another neutron configuration:
$\{\nu 9/2^{-}[734] \otimes \nu 7/2^{+}[624]\}$,
the most favorable to form \mbox{$K$-isomer}. Its energy is two times smaller
 than reported in the experiment.

Theoretical predictions for two-quasiparticle states in $^{252}$No, depending
 on their mutual combinations, give excitation energies in the range
$\langle 0.9 \div1.4 \rangle~MeV$.
Predictions for the lowest $E^{*}$ usually come from the Woods-Saxon model
 with the Lipkin-Nogami version of including pair correlations
 \cite{Kondev2007}, while the highest excitations are predicted e.g. within
 the quasiparticle phonon model \cite{Jolos2011}, namely $ E^{*}=1.300~MeV$ for
$\nu^2 8^{-}= \{\nu 9/2^{-}[734] \otimes \nu 7/2^{+}[624]\}$
and $E^{*}=1.336~MeV$ for $\pi^2 8^{-}=\{\pi7/2^{-}[514] \otimes \pi9/2^{+}[624]\}$.

Next comparison can be done for $^{250}$No for which quite recently stability
of the \mbox{high-$K$} isomer was investigated by Kallunkathariyil et al.
\cite{Kallunkathariyil2020}. In the previous study \cite{Peterson2006} a
 tentative assignment of the isomeric state was:
$K^{\pi} = \nu^2 6^{+}$, with the following neutron components:
$\{\nu5/2^{+}[622] \otimes \nu 7/2^{+}[624]\}$. The experimental energy is
 not known.
Our calculations fully support such assignment as this configuration
is clearly low laying: $E^{*} \approx 0.6~MeV$
in the case of our already weak pairing interaction within the BCS formalism.
 Higher states (with energies in $MeV$) are collected in Table \ref{250No}.
\begin{table}[h!]
\caption{Specification of excited states in $^{250}No$} \label{250No}
\begin{tabular}{|c|c|c|}
\hline
$  \pi / \nu~K^{\pi} $ & Configuration & $E^{*}$ (MeV)  \\
\hline
$\nu^2 6^{+}$ & $\{\nu5/2^{+}[622] \otimes \nu7/2^{+}[624]\}$ & 0.57  \\
$\nu^2 7^{-}$ & $\{\nu5/2^{+}[622] \otimes \pi9/2^{-}[734]\}$ & 0.76  \\
$\nu^2 8^{-}$ & $\{\nu7/2^{+}[624] \otimes \pi9/2^{-}[734]\}$ & 0.97  \\
$\nu^2 4^{+}$ & $\{\nu7/2^{+}[624] \otimes \pi1/2^{+}[631]\}$ & 1.10  \\
$\nu^2 7^{-}$ & $\{\nu7/2^{-}[743] \otimes \pi7/2^{+}[624]\}$ & 1.21  \\
$\nu^2 5^{-}$ & $\{\nu1/2^{+}[631] \otimes \pi9/2^{-}[734]\}$ & 1.24  \\
$\nu^2 8^{+}$ & $\{\nu7/2^{-}[743] \otimes \pi9/2^{-}[734]\}$ & 1.34  \\
\hline
\end{tabular}
\end{table}
Also the modified two-center shell model predicts this $K^{\pi} =
\nu^2 6^{+}$ state as the most likely \mbox{$K$-isomer} ($E^{*}$ about $1.2~MeV$).
Liu et al. \cite{Liu2014}, selects this configuration as the favored one for
the two-quasipartcle isomer in $^{250}$No as well
 (with energy $E^{*}=0.83~MeV$).

Finally we would like to compare our data to the recent measurement in
 $^{254}$Rf.
Using digital electronic, under vacuum conditions at the Fragment Mass
 Analyser at Argonne, two signals suspected of 2-qp configurations:
 $\pi^2 8^{-} = \{\nu 9/2^{-}[734] \otimes \nu 7/2^{+}[624]\}$
or $\pi^2 8^{-}=\{\pi7/2^{-}[514] \otimes \pi9/2^{+}[624]\}$ were registered.
 The 4-qp isomeric state, likely: $\{\pi^2 8^{-} \otimes \nu^2 8^{-} \}$,
 involving one of those, has been found by David et al. \cite{David2015}.
 These are extremely low-lying states according to our calculations,
 see Fig.~\ref{Ex104}.
 Possible underestimate of $E^{*}$ in our calculations has its source in the
  BCS treatment of pairing: the blocking procedure within this method
 induces a too large reduction in the pairing gap for multi-quasiparticle
 states and causes an underestimate in their excitation energies.
 Therefore our $E^{*}$ may tend to be lower than the corresponding
 experimental excitations.
 One possibility to avoid such deficiency would be to assume a stronger
 pairing interaction for considered multi-qp configurations.
 However, because such modification will not affect the sizes and shapes of
 studied multi-quasiparticle configurations and (according to our tests)
 only systematically increases calculated $E^{*}$,
 we decided to keep all our BCS parameters without any adjustment.

\subsection{Role of the high rank deformations}

\begin{figure}[h]
\centerline{\includegraphics[scale=0.7]{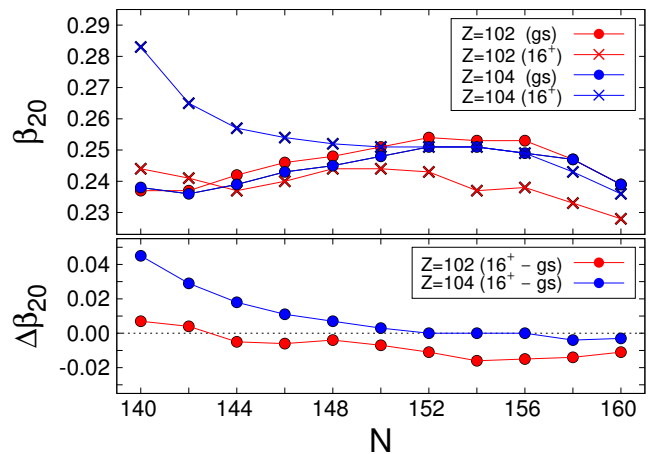}}
\caption{{\protect Upper panel: Values of the deformation parameter $\beta_{20}$ in
the ground (gs) and excited 4-qp states
$\pi^2 \nu^2 16^{+}=\{\pi^2 8^{-} \otimes \: \nu^2 8^{-}\}$
($16^{+}$), for No and Rf isotopes (red and blue color, respectively).
Bottom panel: Differences between $\beta_{20}$ values in the $16^+$
and gs states for two isotopic chains: \mbox{$Z=102$} and \mbox{$Z=104$} (red and blue color, respectively).
}}
\label{B20_changes}
\end{figure}

\begin{figure}[h]
\centerline{\includegraphics[scale=0.7]{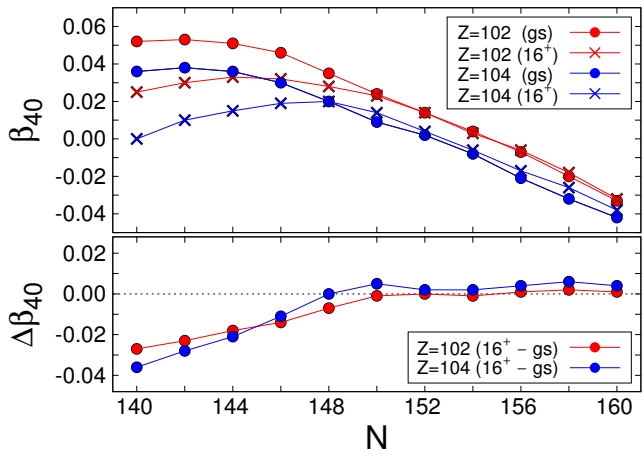}}
\caption{{\protect The same as in Fig.~\ref{B20_changes},
but for the deformation parameter $\beta_{40}$.}}
\label{B40_changes}
\end{figure}

\begin{figure}[h]
\centerline{\includegraphics[scale=0.7]{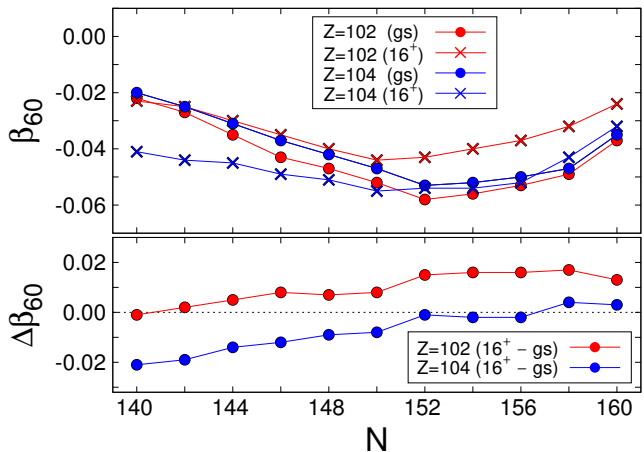}}
\caption{{\protect The same as in Fig.~\ref{B20_changes},
but for the deformation parameter $\beta_{60}$.}}
\label{B60_changes}
\end{figure}

\begin{figure}[h]
\centerline{\includegraphics[scale=0.7]{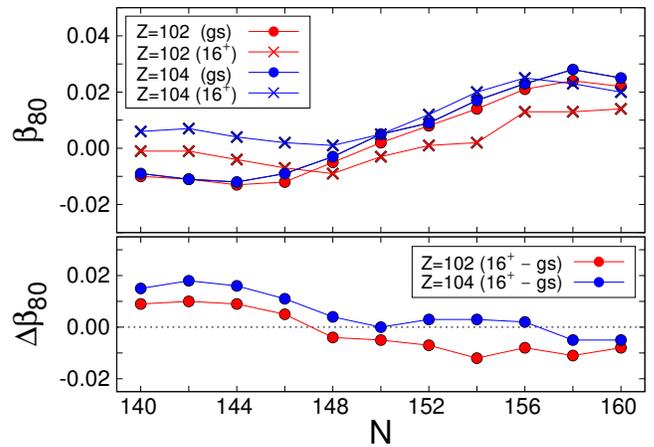}}
\caption{{\protect The same as in Fig.~\ref{B20_changes},
but for the deformation parameter $\beta_{80}$.}}
\label{B80_changes}
\end{figure}

The MM method used here allows to examine step by step
the role of deformation parameters of the higher order.
Patyk and Sobiczewski in \cite{Patyk19911,Patyk19912} noted a wider shell
gap around $Z=100$ and $N=150$ after the inclusion of $\beta_{60}$ in their definition of the nuclear radius,
which gave them much better agreement with existing experimental data.
Recently, an effect of the deformation $\beta_{60}$ on \mbox{high-$K$} isomer
properties in superheavy nuclei has been discussed by Liu et al.~\cite{Liu2011}.
Here, we use even one more deformation -  $\beta_{80}$.
In this subsection we would like to discuss changes in
deformation parameters $\beta_{20},\beta_{40},\beta_{60},\beta_{80}$,
along isotopic chains, both for the ground states as well for \mbox{high-$K$}
 configurations.
As an example of such an excited configuration we will consider
 the mentioned ealier:
$\pi^2 \nu^2 16^{+}=\{\pi^2 8^{-} \otimes \: \nu^2 8^{-}\}$,
build on the following two-quasiproton:
$\pi^2 8^{-}=\{7/2^{-}[514] \otimes 9/2^{+}[624]\}$
and two-quasineutron:
$\nu^2 8^{-}=\{7/2^{+}[624] \otimes 9/2^{-}[734]\}$
excitations. For simplicity, this 4-qp state is indicated in
Figs.~\ref{B20_changes}-\ref{B80_changes} as: $16^{+}$.

 The quadrupole deformation ($\beta_{20}$) effect on energy
 is the largest - Fig.~\ref{B20_changes}.
One can see on the top panel of this figure that No nuclei
(red filled circles) are
in most cases slightly more quadrupole deformed at the ground
states than the Rf nuclei (blue filled circles).
First and foremost, one can see systematically different behavior
of this parameter for the two tested isotopic chains in excited states.
Except two cases: $^{242}$No and $^{244}$No, in all considered nobelium nuclei
the quadrupole deformation is smaller in excited states
compared to the ground states.
However such differences, namely: $\Delta \beta_{20} = \beta^{ex}_{20}-\beta^{gs}_{20}$,
are not large as shown on the bottom panel of
Fig.~\ref{B20_changes} (red circles), where it can be seen that they grow
insignificantly for heavier isotopes. For Rf isotopes lighter than
 $^{254}$Rf, the differences (blue
circles on the bottom panel in Fig.~\ref{B20_changes}), are clearly
 increasing as the number of neutrons decreases. In the extreme case, the
 deformation difference of considered excited
state $\pi^2 \nu^2 16^+$ with respect to the gs reaches up to 20 percent.
Quadrupole deformations of the excited state for heavier Rf nuclei are
 actually the same as those at the ground state.

The isotopic behavior of hexadecapole deformation for No and Rf isotopes is shown in Fig.~\ref{B40_changes}.
Also this time the ground states values of this shape variable are systematically and clearly higher
for No than for Rf. In contrast to what we observed for the quadrupole
in both chains the course of variation $\Delta \beta_{40} = \beta^{ex}_{40}-\beta^{gs}_{40}$
is now similar i.e.: their values for excited 4-qp configurations
are negative for lighter nuclear systems and very close to zero for the heavier ones ($N>148$).
This is reflected on bottom panel in the Fig.~\ref{B40_changes}.
We would like to emphasize that relative changes of this hexadecapole deformation for considered 4-qp state
are really enormous for lighter isotopes.

Yet another behavior is revealed by $\beta_{60}$ parameter what has been shown in Fig.~\ref{B60_changes}.
In the sense of the absolute value of this variable nobelium nuclei are almost always much more deformed
at the ground states than at the considered $\pi^2 \nu^2 16^+$ excited state (top panel of Fig.~\ref{B60_changes}).
In rutherfordium nuclei the situation is a little different, namely: for $N<152$ excited nuclei are
visibly less deformed in this direction compared to the ground states. For $N\ge152$ the $\beta_{60}$ deformation
of considered high-$K$ configuration are quite similar to the corresponding ground state configurations.
The differences, $\Delta \beta_{60} = \beta^{ex}_{60}-\beta^{gs}_{60}$, of this parameter
in the $N$ chains are shown on the bottom panel of Fig.~\ref{B60_changes}.

For completeness, we also plotted the
deformations $\beta_{80}^{gs}$ and $\beta_{80}^{ex}$ as the function of the
neutron number $N$ (top panel of Fig.~\ref{B80_changes}).
Of course, these changes are rather not as important as the previous ones.
However, it can be seen that the values of this deformation parameter for both
 chains are almost the same in ground states. Also the isotopic variation
 of $\Delta \beta_{80}$ is similar (the bottom panel of Fig.~\ref{B80_changes}).
\begin{figure}[h]
\centerline{\includegraphics[scale=0.5]{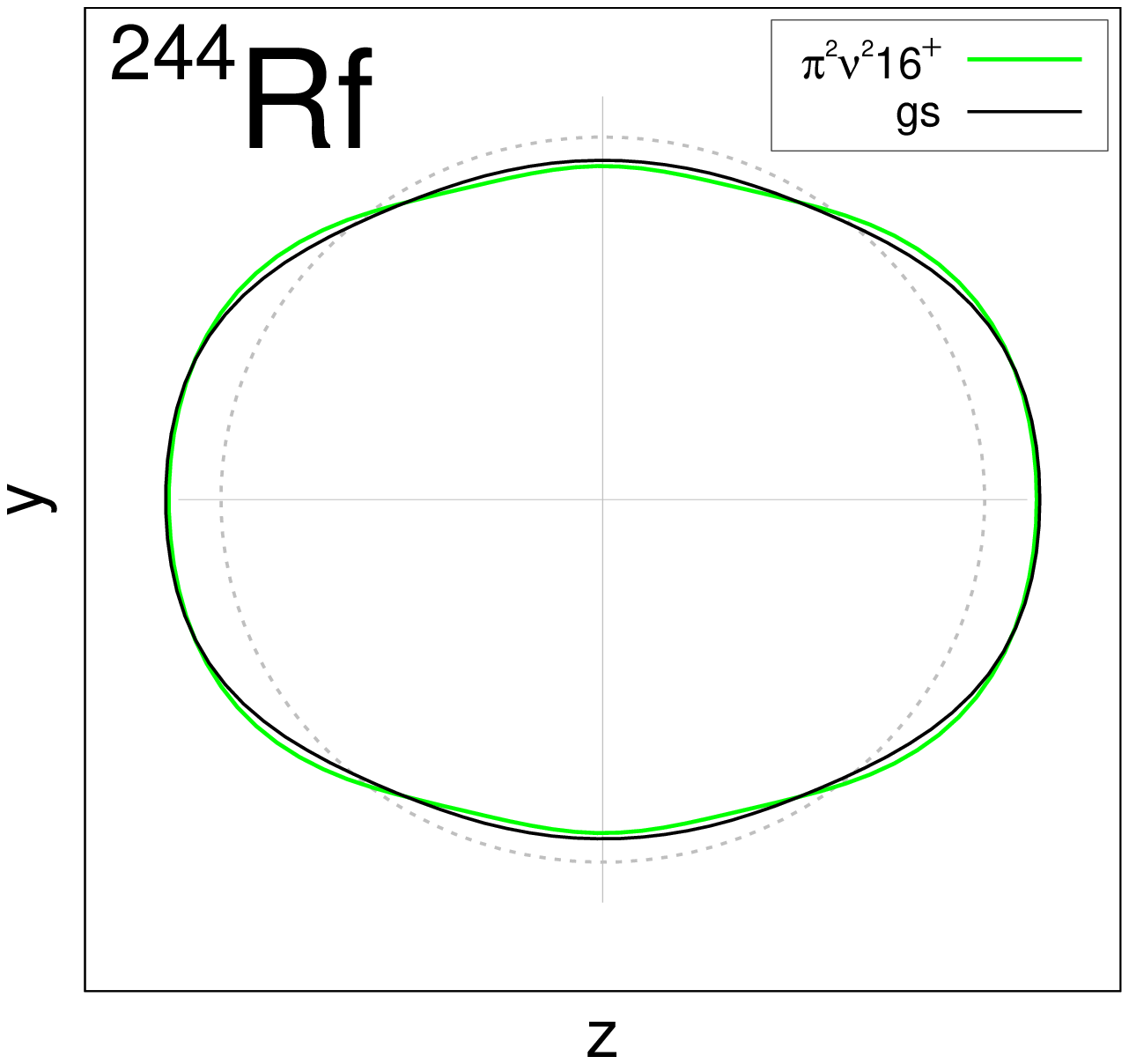}}
\centerline{\includegraphics[scale=0.5]{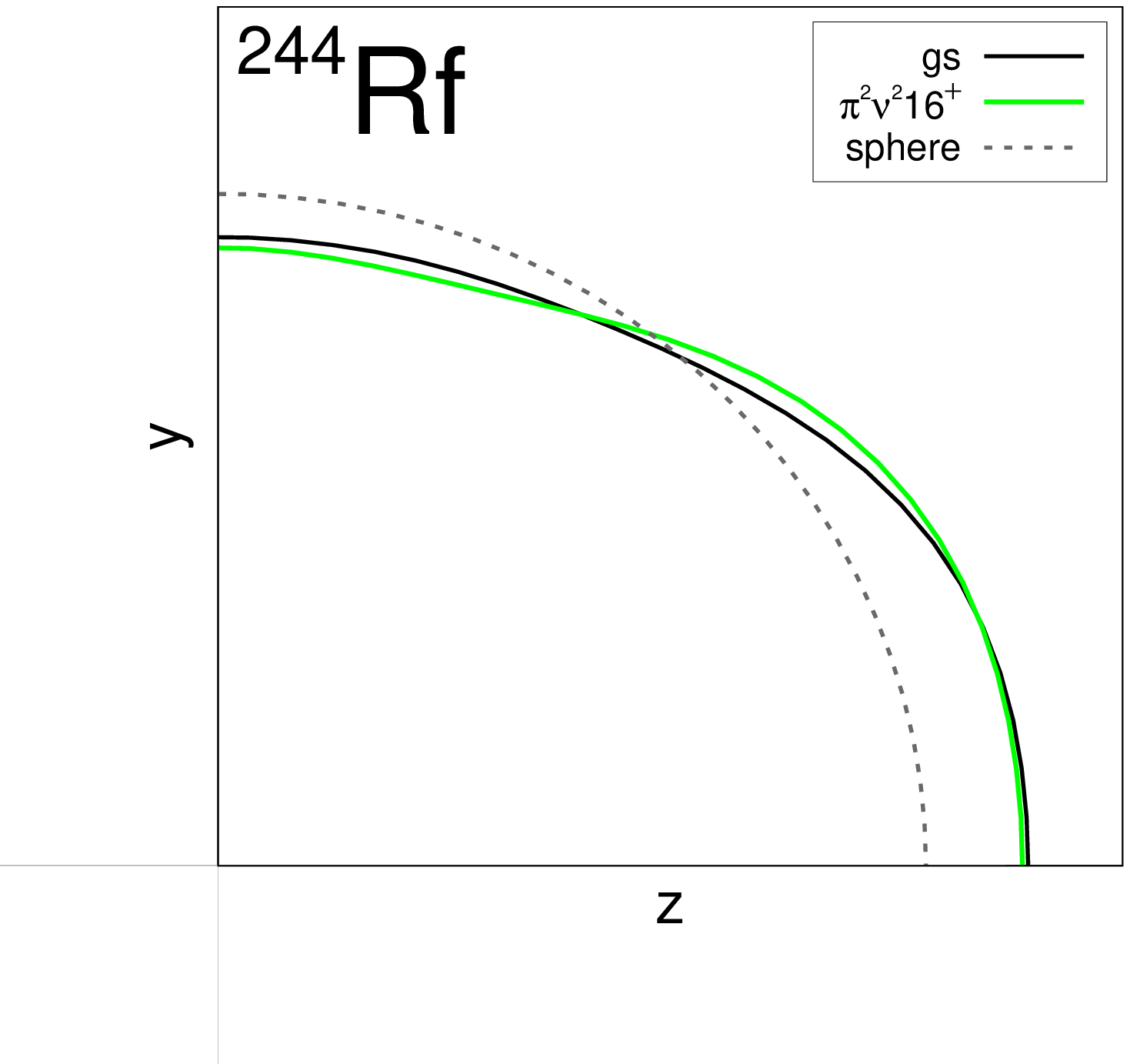}}
\caption{{\protect Cut of a sample shape in the $y-z$ plane for $^{244}$Rf.
Green line - calculation for 4-qp excited state ($\pi^2 \nu^2 16^{+}$);
black line - calculation for ground state (gs).}}
\label{244Rfshape}
\end{figure}

Finally, in Fig.~\ref{244Rfshape}, the shapes for $^{244}$Rf nucleus corresponding
to its ground (on black) as well to excited \mbox{high-$K$}
state $\pi^2 \nu^2 16^{+}$ (green color), are shown. This is exactly the same
4-qp configuration as considered previously in Figs. \ref{B20_changes}-\ref{B80_changes}.
In the scale of the obtained values of deformations, the differences
between this two shapes are not very significant. However, in the scale of energy these differences
are very clear and, for example in this particular nucleus, the energy difference (i.e. excitation energy $E^{*}$)
is about $3~MeV$.

\subsection{Electric quadrupole moments}

A deviation of the proton distribution from sphericity
 is characterized by the electric quadrupole moment.
 Assuming a uniform distribution of the charge $Ze$ within a sharp, deformed
 nuclear surface, i.e. neglecting the diffuseness of the nuclear surface,
  one obtains the intrinsic electric quadruple moment:
\begin{equation}
\label{Q20def}
Q_{20}^{ex/gs} = \frac{3Ze}{4 \pi  r^{3}_{0}} \int^{R(\theta)} _{0}    r^{2}(\theta)  P_{20}(\theta) d^{3}r.
\end{equation}
 In the leading order, $Q_{20}$ is proportional to $\beta_{20}$, but
 other deformations are also included in the above formula.
 Within the MM model, we disregard possible differences between
 the proton and neutron deformations.
 The moment (\ref{Q20def}) is calculated in the intrinsic frame, while
 measurements are performed in the laboratory frame. The usually assumed
 transformation between the two neglects the Coriolis effects. This is
  substantiated for \mbox{high-$K$} states in well deformed nuclei,
 in which $K$ is good quantum number.
 In this strong coupling limit one can transform intrinsic quadrupole moment
 to the laboratory frame in a simple way:
\begin{equation}
\label{Q20exp}
Q_{20}^{ex/gs} (exp) = \frac{3K^2 - I(I+1)}{(I+1)(2I+3)} Q_{20}^{ex/gs} ,
\end{equation}
where $K$ is the projection of $I$ on the (intrinsic) symmetry axis.

\begin{figure}[h]
\centerline{\includegraphics[scale=0.7]{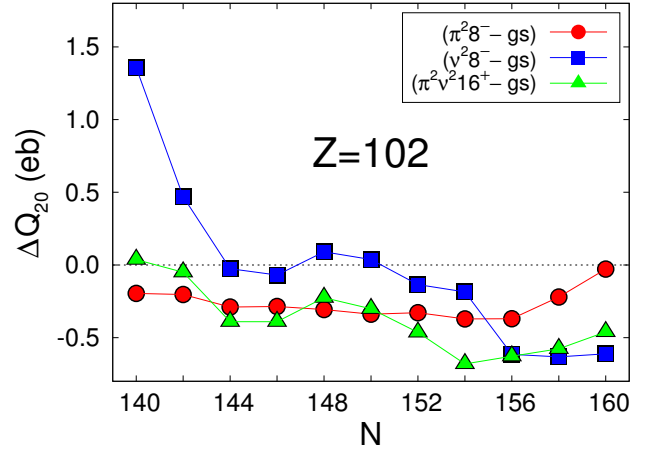}}
\caption{{\protect The changes of the electric quadrupole moments
in case of selected 2- and 4-qp configurations in No isotopes.
Configurations:
$\pi^2 8^{-}=\{\pi7/2^{-}[514] \otimes \pi9/2^{+}[624]\}$,
$\nu^2 8^{-}=\{\nu7/2^{+}[624] \otimes \nu9/2^{-}[734]\}$,
$\pi^2 \nu^2 16^{+}=\{\pi^2 8^{-} \otimes \: \nu^2 8^{-}\}$.
The lines are drown to guide an eye.}}
\label{Q20No}
\end{figure}

\begin{figure}[h]
\centerline{\includegraphics[scale=0.7]{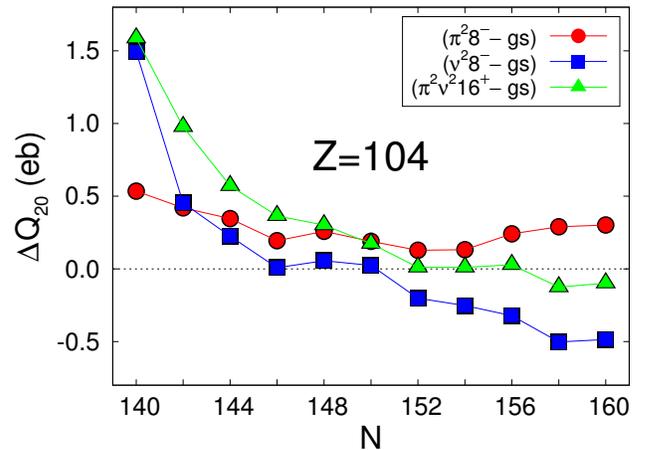}}
\caption{{\protect  The same as in Fig.~\ref{Q20No}, but in Rf isotopes.}}
\label{Q20Rf}
\end{figure}

The calculated differences in the quadrupole moment between a given
 \mbox{high-$K$} state ($Q_{20}^{ex}$) and the ground state ($Q_{20}^{gs}$):
\begin{equation}
\Delta Q_{20} = Q_{20}^{ex} - Q_{20}^{gs},
\end{equation}
are shown in Fig.~\ref{Q20No} and Fig.~\ref{Q20Rf}, for No and Rf isotopes, respectively.
Here and in the next subsection, the chosen states are marked in the same way:
$\pi^2 8^{-}$ by red dots, $\nu^2 8^{-}$ by blue squares, and
 $\pi^2 \nu^2 16^{+}$ by green triangles.
The dependence of $\Delta Q_{20}$ vs $N$ of excited neutron configuration is
 very similar in both elements, unlike for the proton configuration.
  Quadrupole moments of excited proton configurations
 are bigger than in the ground state in Rf nuclei, while
 the difference is  negative in No nuclei.
 As expected, with the fixed number of protons, $\Delta Q_{20}$ for
 $\pi^2 8^{-}$ states does not show a strong dependence on $N$.
 In the lightest isotopes, the two-quasineutron state has the largest
 difference $\Delta Q_{20}$, while $\Delta Q_{20}$ turns negative for the
 heaviest shown isotopes.
However, the most spectacular differences $\Delta Q_{20}$ we observe for
 the considered four-quasiparticle state $\pi^2 \nu^2 16^{+}$ (green colour).
 In most of No nuclei, the calculated $Q_{20}$ is smaller in the exited state
 while for Rf nuclei the \mbox{high-$K$} state is predicted to have larger
 $Q_{20}$.
The difference $\Delta Q_{20}$ is the largest for the lightest Rf isotopes,
 especially in $^{244}$Rf.

\subsection{Mean-square charge radii}

 The mean-square charge radius: ${\langle r^{2} \rangle}$ characterizes the
 size of a nucleus. It is extracted from the measured
 atomic hyperfine structure which is determined by analysing atomic
 hyper - fine transitions.
 This structure is very sensitive to even slight changes in
  ${\langle r^{2} \rangle}$, in particular those rasulting from
  shape changes.
 The droplet model - like  expression for ${\langle r^{2} \rangle}$ consists of
 four terms \cite{K-HSchmidt1983,Buchinger1994}:
\begin{equation}
\label{deltardroplet}
 {\langle r^{2} \rangle} = {\langle r^{2} \rangle}_{u} + {\langle r^{2} \rangle}_{r} + 3\sigma^{2} + s_{p}^{2} ,
\end{equation}
where:  ${\langle r^{2} \rangle}_{u}$ is the "{\it uniform}" term; ${\langle r^{2} \rangle}_{r}$ is the "{\it Coulomb redistribution}" term;
$3\sigma^{2}$ comes from the diffuseness of the nuclear surface and $s_{p}^{2}$  is the finite-proton-size contribution.
 Interesting for us will be the difference in ${\langle r^{2} \rangle}$
 between the high-K configuration and the spherical shape
of a given nucleus:
\begin{equation}
\label{deltar2}
\delta {\langle r^{2} \rangle} = {\langle r^{2} \rangle}_{def} - {\langle r^{2} \rangle}_{sph},
\end{equation}
where ${\langle r^{2} \rangle}_{def}$ corresponds to deformation. According to
 this definition, $\delta {\langle r^{2} \rangle}$ can be understood as a
 measure of deviation from the spherical shape, characterized in
 (\ref{deltar2}) by ${\langle r^{2} \rangle}_{sph}$.

We are going to discuss in this section isotopic dependance of
 $\delta {\langle r^{2} \rangle}$ not only for the ground states, but
rather for selected multi-qp excited \mbox{high-$K$} configurations i
 n realation to the ground states.
 A difference in $\delta {\langle r^{2} \rangle}$ between ground and excited
 states is:
\begin{equation}
\label{difr2}
\Delta \delta \langle r^{2} \rangle = \delta \langle r^{2} \rangle ^{ex} - \delta \langle r^{2} \rangle^{gs},
\end{equation}

In this difference, the last two components of Eq. \ref{deltardroplet} cancel.
The Coulomb redistribution term ${\langle r^{2} \rangle}_{r}$, which
can be evaluated according to formula (23) in \cite{K-HSchmidt1983} is already
 a small quantity compared to ${\langle r^{2} \rangle}_{u}$.
It seems that the influence of the Coulomb redistribution effect on the
 difference (Eq. \ref{difr2}) may be neglected.

Thus, using all obtained deformations $\beta_{\lambda 0}$ listed in (\ref{radius})
one can calculate $\delta {\langle r^{2} \rangle} \simeq \delta {\langle r^{2} \rangle _{u}}$ (in a quite general way, see \cite{Scheidenberger})
separately for deformed ground states (gs) and excited states (ex), as:%
\begin{equation}
\label{r2}
 {\delta {\langle r^{2} \rangle}^{ex/gs}}  =  {\langle c^2 R^{2}_0 \rangle}^{ex/gs} \cdot [ 1+\frac{5}{4\pi}\sum_{\lambda=2,4,6,8}{\beta_{\lambda 0}^{2}}^{ex/gs} ].
\end{equation}

Seemingly, the changes of $\delta {\langle r^{2} \rangle}$ are very small,
 usually they are of the order of a few percent, but are measurable.
For example, a relative change in mean-square charge radius due to one
neutron less from the isotope shift can be estimated (in the spherical case),
as:
$\delta {\langle r^{2} \rangle}^{A,A-1}_{sph} / \delta {\langle r^{2} \rangle}_{sph} =1/(3A) $,
see \cite{Scheidenberger}, what means for a very heavy nuclei - considered here (A $\approx$ 250) about 0.1 percent.
 Now the need for a detailed analysis of the deformation parameters made in the previous paragraph becomes quite clear.

\begin{figure}[h]
\centerline{\includegraphics[scale=0.7]{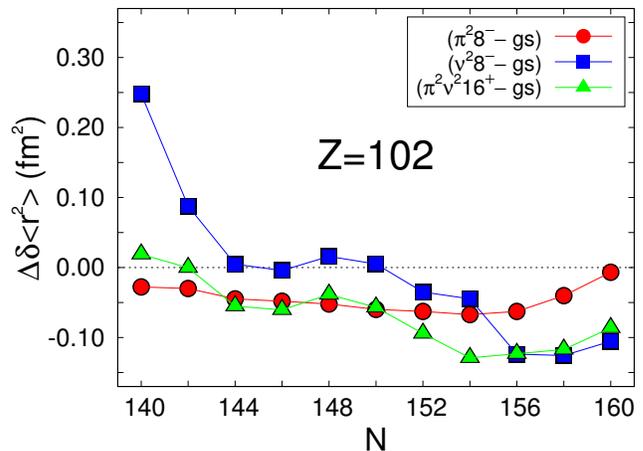}}
\caption{{\protect The changes in mean-square nuclear charge radii
$\delta \langle r^{2} \rangle$ for the selected 2- and 4-qp configurations in No isotopes.
Configurations:
$\pi^2 8^{-}=\{\pi7/2^{-}[514] \otimes \pi9/2^{+}[624]\}$,
$\nu^2 8^{-}=\{\nu7/2^{+}[624] \otimes \nu9/2^{-}[734]\}$,
$\pi^2 \nu^2 16^{+}=\{\pi^2 8^{-} \otimes \: \nu^2 8^{-}\}$.
The lines are drown to guide an eye.}}
\label{dr2No}
\end{figure}

\begin{figure}[h]
\centerline{\includegraphics[scale=0.7]{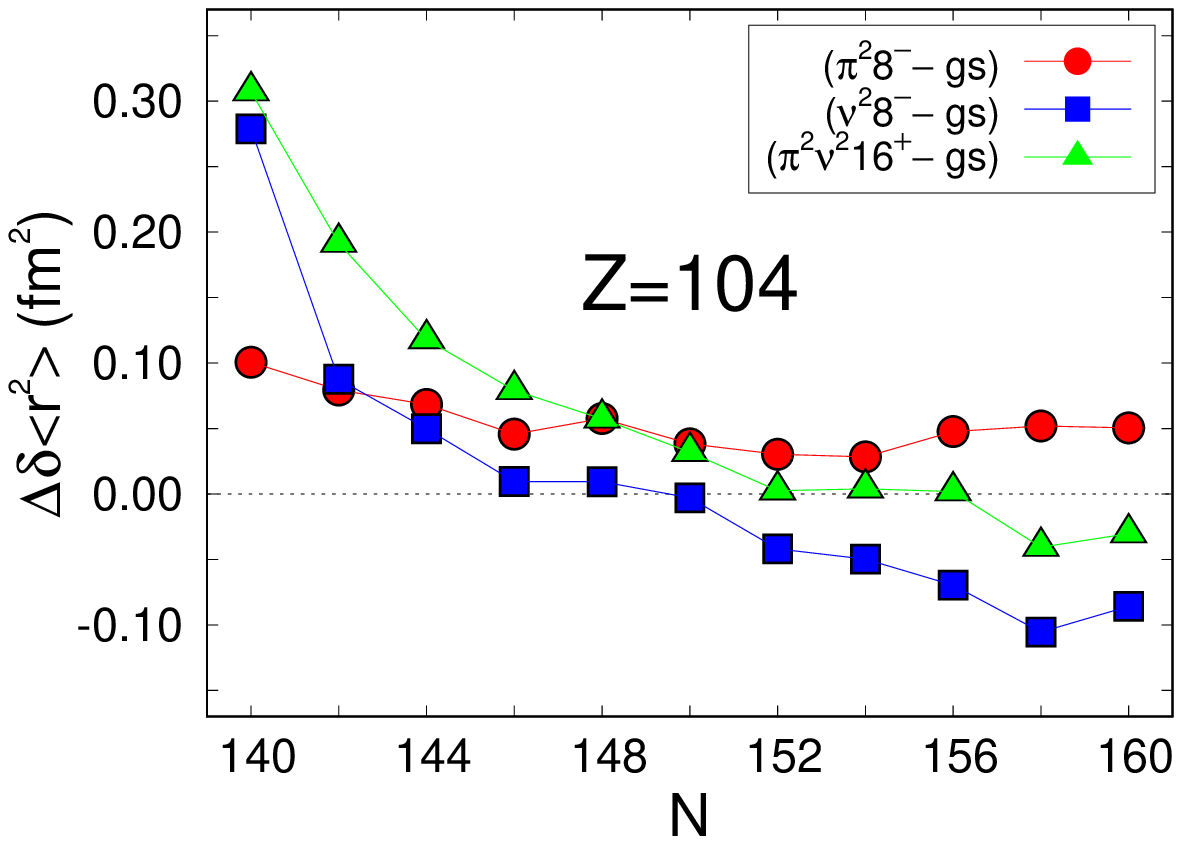}}
\caption{{\protect The same as in Fig.~\ref{dr2No}, but in Rf isotopes.}}
\label{dr2Rf}
\end{figure}

The results based on definition \ref{difr2} are shown in Figs.~\ref{dr2No},~\ref{dr2Rf}.

 One can assume that the size of a given nucleus at the spherical shape for
 both configurations (gs and ex) is the same:
$\langle c^2 R^{2}_0 \rangle^{gs} = \langle c^2 R^{2}_0 \rangle^{ex}$,
and than one can rewrite Eq.~\ref{difr2} in the explicit form:
\begin{equation}
\label{difr3}
\Delta \delta \langle r^{2} \rangle= R^{2}_0 \frac{5}{4\pi}\sum_{\lambda=2,4,6,8}{ \langle \beta^{2}_{\lambda 0}} \rangle ^{ex}-{ \langle \beta^{2}_{\lambda 0}} \rangle^{gs}.
\end{equation}
However, we prefer to discuss numerical
results with the properly calculated volume coefficient $c$ in radius expansion. 

 Looking at Figs.~\ref{dr2No},~\ref{dr2Rf}, one can see that the results for
 charge radii imitate those obtained for quadrupole moments,
for all three states considered here.
In principle, everything that has been said about $\Delta Q_{20}$ vs $N$
 for these states now can be repeated for
 $\Delta \delta \langle r^{2} \rangle$.
In particular, the behavior of both quantities in Rf and No for
 the corresponding \mbox{high-$K$} states is similar.
As before in the case of quadrupole moments, for configuration
 $\pi^2 \nu^2 16^{+}$ the dominant contribution to the observable effect
in  $\Delta \delta \langle r^{2} \rangle$ comes from the two-quasiparicle
 neutron excitation.
One can clearly see that the plot for the neutron configuration is quite
 similar to that for the 4-qp excitation.

Finally, we would like to note that the calculated changes in the size-shape
 quantities are the smallest wherever the \mbox{$K$-isomer} is most
 likely.
For example, $\nu^2 8^{-}=\{\nu7/2^{+}[624] \otimes \nu9/2^{-}[734]\}$ state,
 suspected to be isomeric in $^{252}$No
\cite{Herzberg2001,Sulignano2012,Leppanen2006}
or/and in $^{254}$Rf \cite{David2015,Khuyagbaatar2020}, that is supported
 here by Figs.: \ref{Ex102}, \ref{Ex104}, \ref{Ex150}, does not show almost
 any change of shape relative to the g.s.
It is not surprising, as isomers occur where the excitation energy is
 relatively low.
This means that the metastable minimum is slightly above the global minimum.
The proximity of these minima means that the deformations for
 excited states are similar to those for the gs. Consequently, the
 quantities such as $Q_{20}$ and $\delta \langle r^{2} \rangle$  do not differ
 much from those for the g.s. In a sense, the lack of change of the nuclear
 shape in isomers may be just an indicator of their isomeric nature.

\section{CONCLUSIONS}

We have shown on selected examples how the size of the very heavy nuclear system changes if one exited it.
Such a studies seems are of particular importance for nuclei in which such states may have
an isomeric character as they may be relatively long lived.
After discussing the quality of our results with existing experimental data for excitation energy
we have indicated from isotopic chains such candidates.
They are particularly clearly visible for $N=150$, i.e. for: $^{252}$No and $^{254}$Rf,
in all cases of considered multi-quasiparticle configurations.

Calculated charge radii and quadrupole moments for such exited states are compared with
results for the ground states.
The most pronounced difference in the behavior between nobelium and ratherfordium
isotopes in excited states was observed for the considered four-quasiparticle configuration
$\pi^2 \nu^2 16^{+}$ build on two-quasiproton:
$\{\pi7/2^{-}[514] \otimes \pi9/2^{+}[624]\}$
and two-quasineutron:
$\{\nu7/2^{+}[624] \otimes \nu9/2^{-}[734]\}$
excitations.

Also for those 4-qp configurations electric qudrupole momoments and charge radii
differ most significantly with the values of ground states.

Admittedly, for some \mbox{high-$K$} states, the shape changes found are significant,
but just for those suspected of isomerism they are rather negligible
and this lack of shape change may just indicate on the isometric nature of the studied nuclear system.

The leading role of quadrupole deformation was shown and the impact of other shape parameters was discussed in details.

Since, the investigated No and Rf isotopes are not too far away from today's possibilities
of experimental laser spectroscopy techniques
above hypothesis and predictions for \mbox{high-$K$} states may be verify soon.

\section*{ACKNOWLEDGEMENTS}

M.~K. was co-financed by the National Science Centre under Contract No. UMO-2013/08/M/ST2/00257  (LEA COPIGAL).

\end{document}